\documentclass[10pt,conference]{IEEEtran}
\IEEEoverridecommandlockouts

\usepackage{amsfonts}
\usepackage{cite}
\usepackage{amsmath,amssymb,amsfonts}
\usepackage{algorithmic}
\usepackage{graphicx}
\usepackage{textcomp}
\usepackage{xcolor}
\usepackage{multirow}
\usepackage{float}
\usepackage{amssymb}
\usepackage{caption}
\usepackage{subcaption}
\usepackage{algorithm}
\usepackage{hyperref}

\newcommand{\toolname}{DialogAgent}

\def\BibTeX{{\rm B\kern-.05em{\sc i\kern-.025em b}\kern-.08em
    T\kern-.1667em\lower.7ex\hbox{E}\kern-.125emX}}

\begin{document}

%
\title{DialogAgent: An Auto-engagement Agent for Code Question Answering Data Production}

%

\makeatletter
\newcommand{\linebreakand}{%
  \end{@IEEEauthorhalign}
  \hfill\mbox{}\par
  \mbox{}\hfill\begin{@IEEEauthorhalign}
}
\makeatother

\author{\IEEEauthorblockN{Xiaoyun Liang}
\IEEEauthorblockA{ByteDance \\
Shenzhen, China \\
liangxiaoyun.duo@bytedance.com}
\and
\IEEEauthorblockN{Jingyi Ren}
\IEEEauthorblockA{ByteDance \\
Shenzhen, China \\
jingyi.422@bytedance.com}
\and
\IEEEauthorblockN{Jiayi Qi}
\IEEEauthorblockA{ByteDance \\
Beijing, China \\
qijiayi@bytedance.com}
    \linebreakand
\IEEEauthorblockN{Chao Peng\textsuperscript{*}}
\IEEEauthorblockA{ByteDance \\
Beijing, China \\
pengchao.x@bytedance.com}
\and
\IEEEauthorblockN{Bo Jiang}
\IEEEauthorblockA{ByteDance \\
Shenzhen, China \\
jiangbo.jacob@bytedance.com}
}


\maketitle

\begingroup\renewcommand\thefootnote{*}
\footnotetext{Corresponding author.}
\endgroup

\begin{abstract}
Large Language Models (LLMs) have become increasingly integral to enhancing developer productivity, particularly in code generation, comprehension, and repair tasks.
However, fine-tuning these models with high-quality, real-world data is challenging due to privacy concerns and the lack of accessible, labeled datasets.
In this paper, we present \toolname{}, an automated tool for generating synthetic training data that closely mimics real developer interactions within Integrated Development Environments (IDEs).
\toolname{} enables the production of diverse, high-fidelity query-response pairs by simulating multi-turn dialogues and contextual behaviors observed in real-world programming scenarios.
The tool significantly reduces the reliance on manual data generation, increasing efficiency by 4.8 times compared to traditional methods.
Our experiments and online deployment demonstrate substantial improvements in model performance for code-related question-answering tasks: the acceptance rate of responses generated by our in-house model is improved by 33\%, after training on synthesized data generated by \toolname.
\end{abstract}

\begin{IEEEkeywords}
    Agent, Large Language Models, Code Question Answering
\end{IEEEkeywords}

\section{Introduction}
\label{sec:introduction}
With the rise of Large Language Models (LLMs), a new generation of LLM-based programming assistant tools has emerged, providing developers with powerful capabilities such as code completion~\cite{cursor, copilot, marscode, codeium}, debugging~\cite{liu2024marscode, zhang2024autocoderover, xia2024agentless}, and natural language-based code understanding~\cite{macneil2023experiences}.
These tools have become indispensable for enhancing developer productivity, offering support in real-time within Integrated Development Environments (IDEs) and during various stages of software development.

However, deploying these programming assistants in real-world scenarios presents significant challenges.
The security and privacy of user data are of paramount importance, particularly when dealing with proprietary code and sensitive information.
This makes it impractical to rely on closed-source commercial APIs, which may expose sensitive data to external services.
Additionally, the demand for faster inference speeds further constrains the deployment of large LLMs, as models must balance computational efficiency with parameter size to ensure usability in local or private environments.
As a result, there is a growing need for personalized smaller-sized LLMs that can be optimized for specific user and business contexts, all while adhering to strict privacy policies.

A key challenge in developing such personalized LLMs lies in the training data~\cite{dong2023abilities}.
While real-world data is invaluable for fine-tuning LLMs, its use in training is often restricted by user agreements and internal security policies.
This makes it impossible to leverage real user data in scenarios involving LLM-based programming assistants.
Thus, generating high-quality synthetic training data that closely mimics real-world development environments becomes crucial for training models that can operate effectively in these contexts.
The synthetic data must reflect the diversity of real-world interactions while allowing precise control over data categories such as business use cases, user perspectives, and code problem scenarios.

To address these challenges, we have developed \toolname{}, a tool designed to efficiently generate synthetic training data that simulates real IDE environments and developer interactions.
Our tool categorizes the data based on critical dimensions such as business scenarios, user intent, reference regions for answers, and the nature of the code-related problems being addressed.
This enables the construction of targeted, high-fidelity synthetic datasets that are well-suited for fine-tuning LLMs in a secure and controlled manner.

Additionally, we introduce and compare manual and automated evaluation standards to assess the effectiveness of the generated data.
These evaluations are tailored to business-specific outcomes, ensuring that the synthetic data not only meets general quality standards but also aligns with practical development needs.
Experimental results demonstrate that \toolname{} is able to generate questions reflecting real-world user queries and high-quality answers compared to human annotators.
In addition, after training on the data produced by \toolname, the acceptance rate of our internal is increased by 33\% which is used by our developers for daily code-related questions answering.



In this paper, we aim to demonstrate the potential of synthetic data generation in overcoming the limitations of real data usage while maintaining high relevance to real-world development scenarios.
Our findings contribute to the ongoing advancement of LLMs in software development, particularly in enhancing their effectiveness within secure, personalized programming assistants.

\section{Background and Related Work}
\label{sec:background}
\subsection{Large Language Models}

Large language models (LLMs) are powerful pre-trained models designed to process and generate natural language.
These models undergo two key training phases: first, they are pre-trained on vast amounts of text in an unsupervised manner, learning general linguistic patterns and structures; then, they are fine-tuned for specific tasks to optimize their performance on particular applications.

LLMs used for code-related tasks can generally be grouped into three categories based on their architecture: encoder-only models, decoder-only models, and encoder-decoder models~\cite{tian2022learning}.
These models, often built on the transformer architecture, are known for their superior ability to learn from large datasets and scale effectively to handle complex tasks.

\begin{itemize}
   \item \textbf{Encoder-only models}, such as CodeBERT~\cite{feng2020codebert}, rely on a bidirectional transformer encoder and attention mechanisms to learn vectorized embeddings of input code sequences. These models are particularly suited for tasks that require understanding and representation of code, such as code classification, code clone detection, and code search, where generation is not the primary objective.
   \item \textbf{Decoder-only models}, like CodeGen~\cite{nijkamp2022codegen}, InCoder~\cite{fried2022incoder}, and Codex~\cite{chen2021evaluating}, utilize an autoregressive transformer decoder to generate code sequences. These models excel at open-ended tasks such as code generation, where the goal is to produce new code from a given input prompt, making them ideal for scenarios like autocomplete or writing code from natural language descriptions.
   \item \textbf{Encoder-decoder models}, such as LEAM~\cite{tian2022learning}, combine both an encoder and a decoder, making them versatile for tasks that involve both understanding and generating code. These models are effective in a range of applications, including code completion, summarization, and generation, allowing them to handle more complex, multi-stage tasks in development workflows.
\end{itemize}

\subsection{Training Data for Code LLMs}

Regardless of their architectural differences, most LLMs can be fine-tuned with task-specific datasets to improve performance on particular tasks.
As a result, the need for high-quality training data has become more critical than ever~\cite{dong2023abilities, wei2024magicoder, liu2024mftcoder}.
In this paper, we focus on supervised fine-tuning (SFT) on query-response pairs derived from real-world developer interactions, which is promising to enhance the performance of LLMs in code-related tasks.
However, obtaining this data remains a significant challenge.

There have been two main methods of collecting such training data: human annotation and automated data generation~\cite{tan2024large, park2024leveraging, zendel2024enhancing, bhat2023large, yu2024fine, ouyang2022training}.
While human annotation provides high-quality, contextually relevant data, it is a time-consuming and costly process that limits scalability.
On the other hand, existing automated approaches often fall short of capturing the diversity and complexity of real developer interactions, resulting in synthetic datasets that do not fully represent real-world behaviors in integrated development environments (IDEs).
These limitations hinder the effectiveness of LLMs when applied to practical coding scenarios, where diversity in query types, code contexts, and problem-solving approaches is critical for model success.

\subsection{Motivation}

IDE users frequently interact with code completion, explanation, and repair tools embedded within the environment.
These interactions often involve multi-step, context-driven queries that span across different programming languages, code files, and even past interactions in the same session.
Therefore, capturing this rich context, 
is paramount for collecting training data, representing diverse real-world scenarios.

\section{DialogAgent}
\label{sec:approach}
\begin{figure*}[h!tbp]
    \centering
    \includegraphics[scale=0.30]{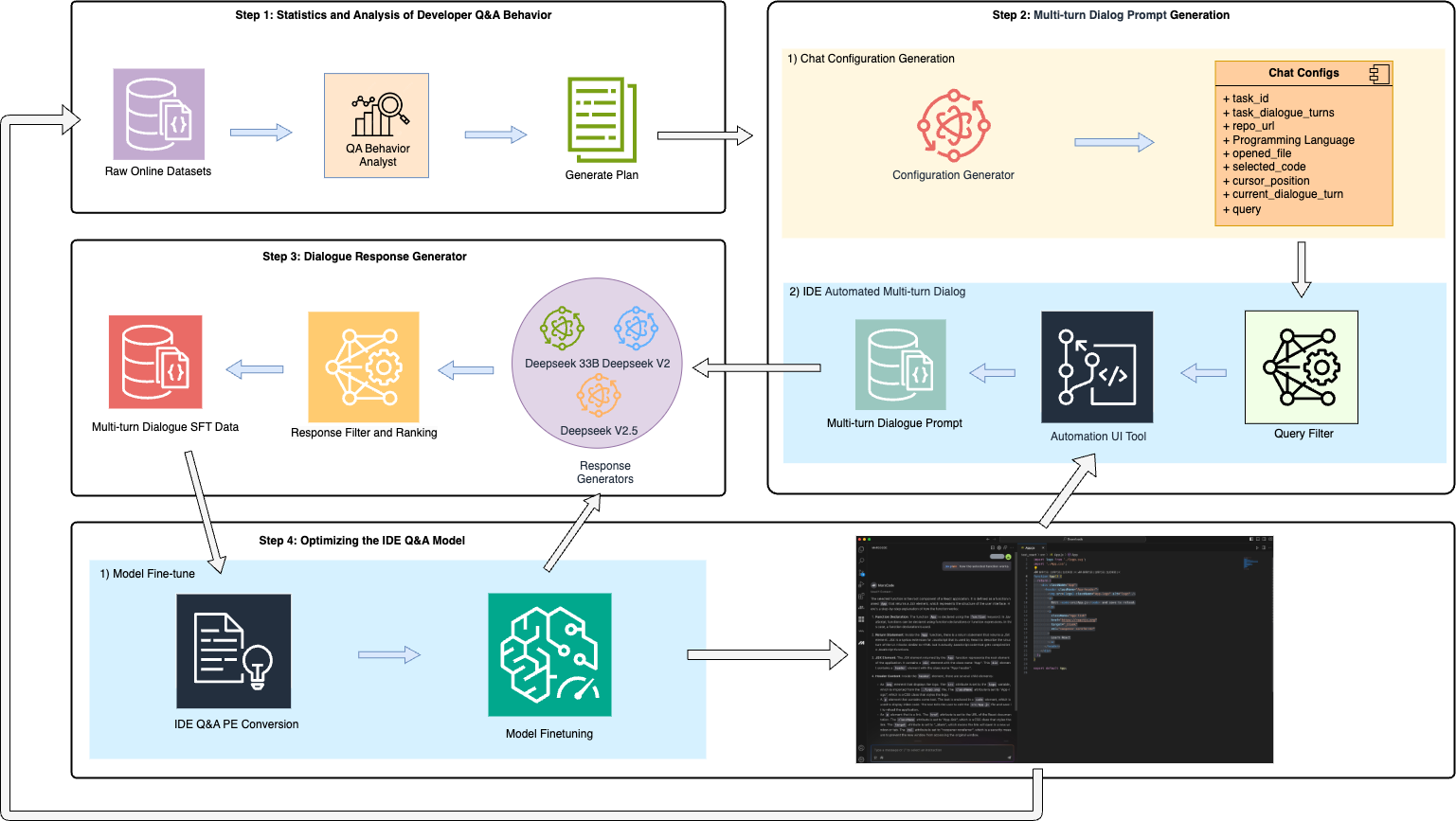}
    \caption{Workflow of \toolname{}}
    \label{fig:Framework}
\end{figure*}

\begin{figure}[h!tbp]
    \centering
    \includegraphics[scale=0.35]{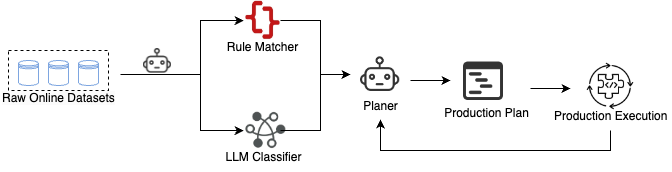}
    \caption{Process of QA-DBA}
    \label{fig:QA-DBA-Framework}
\end{figure}

In this section, we introduce \toolname{}, a novel method designed to automatically generate SFT data at the repository level with a high degree of diversity, quality, and fidelity to real IDE usage.
Figure~\ref{fig:Framework} illustrates the workflow of \toolname{}, which leverages the Q\&A plugin within the VS Code IDE and automates the interaction via a UI automation tool.

Our method begins with a set of existing online user queries, a seed chat model (DeepSeek-Coder-33B~\cite{ds-33b}), and the absence of compliant supervised training data that closely reflects the day-to-day needs of developers.
\toolname{} generates synthetic data by simulating multi-agent interactions within real development environments.
The LLM-based agent acts in three roles: (1) a Q\&A Developer Behavior Analyst (QA-DBA), analyzing and modeling developer behavior in the IDE, and producing a data generation plan;
(2) a chat configuration generator, which constructs detailed conversation configurations (e.g., code context, query, file operations);
and (3) a response generator, which filters and selects high-quality responses from a pool of candidate outputs.

\subsection{QA Developer Behavior Analyst}

The process of QA-DBA is shown in Figure~\ref{fig:QA-DBA-Framework}.
To ensure the generated data is diverse and reflects real online interactions, we first conduct a thorough analysis of developer behaviors using our QA-DBA module.
This agent performs a comprehensive review of Q\&A interactions within IDEs, capturing 10 key behavioral dimensions across more than 40 categories as shown in Table~\ref{table:behavioral categories}, of which the first 7 are classified by the Rule Matcher, and the last 3 are classified by the LLM classifier with the prompt in Figure~\ref{fig:LLM classifier prompt}.
This analysis informs a data production plan, tuned to reflect online developer behavior patterns, as shown in Figure~\ref{fig:LLM planning prompt}.
The QA-DBA framework ensures that the generated training data mirrors real-world developer interactions, maintaining consistency and diversity across different query types, code contexts, and development tasks.
The LLM classifier and planner are both based on the Deepseek-Coder-V2-Instruct~\cite{ds-v2} model.

\begin{table*}[]
    \centering
\footnotesize{
    \caption{QA Developer Behavioral Categories}
    \begin{tabular}{|l|l|}   
    \hline                       
    \textbf{Latitudes} & \textbf{Categories} \\   
    \hline
    Cursor Behavior & no active file, have active file, select a block, select multiple blocks, select a line, select multiple lines \\ 
    \hline
    Triggering Q\&A Method & inline chat, chat view \\ 
    \hline
    Instruction Type & query, quick chat template + query, quick chat template \\ 
    \hline
    Programming Language & python, go, cpp, java, javascript, typescript, etc.\\
    \hline
    System Locale & Chinese, English \\
    \hline
    Dialog Turns & 1, 2, 3, 4, etc.\\
    \hline
    Locale Requirements in Query & different from system locale, same with system locale, no requirement \\
    \hline
    Response Reference Regions & historical dialog, selected code, context, question, error messages, general knowledge.\\
    \hline
    Difficulty Level & elementary, intermediate, advanced, expert \\
    \hline
    Intent & code generation, code editing, code explanation, comment generation, code repair and code general Q\&A\\
    \hline
    \end{tabular}
    \label{table:behavioral categories}
    }
\end{table*}

\begin{figure}
    \centering
    \includegraphics[width=\columnwidth]{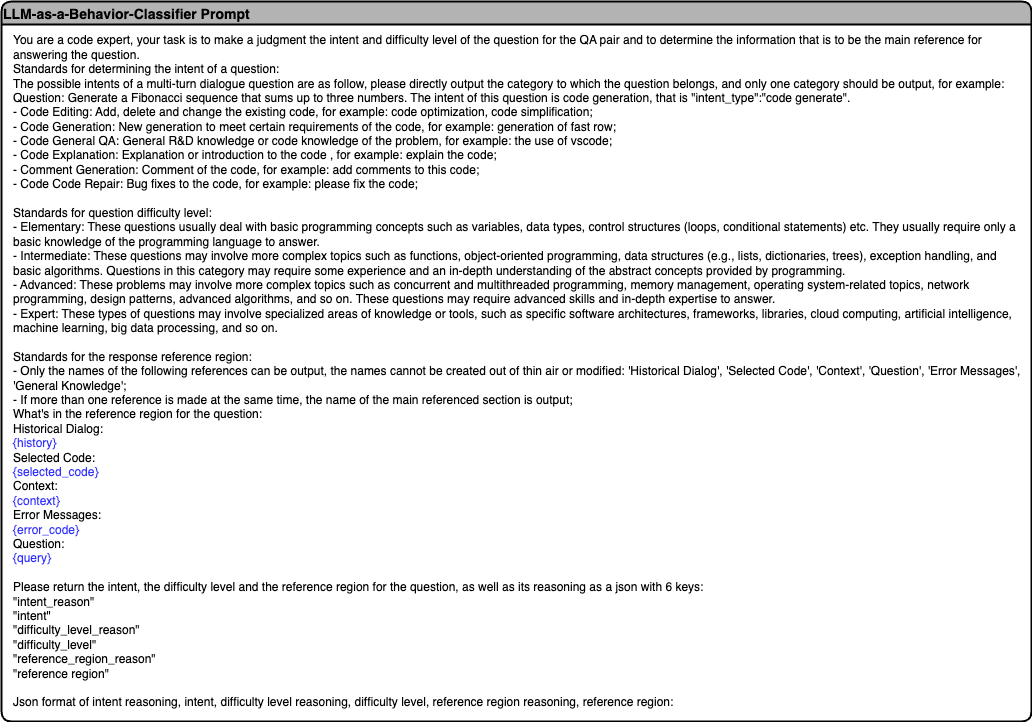}
    \caption{Prompt for Behavior Judgment.}
    \label{fig:LLM classifier prompt}
    \vspace{-10pt}
\end{figure}

\begin{figure}
    \centering
    \includegraphics[width=\columnwidth]{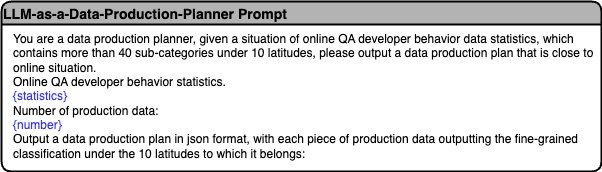}
    \caption{Prompt for Data Production Planning.}
    \label{fig:LLM planning prompt}
    \vspace{-10pt}
\end{figure}
    
\subsection{Chat Configuration Generator}
The chat configuration generator takes as input the interaction context based on developer behavior patterns analyzed by QA-DBA.
These configurations define the repository setup (e.g., cursor position, selected code) and trigger mechanisms (e.g., query type, dialog turn) within the IDE (as illustrated in Figure~\ref{fig:Interact}).
Based on the generated configuration with \textit{Cursor Behavior} and \textit{Programming Language} definitions, an open-source repository from BigCode ~\cite{lozhkov2024starcoder} randomly selected to craft queries.
A sample query generation prompt is illustrated in Figure~\ref{fig:query generating prompt}, which incorporates various attributes such as \textit{Query Locale}, \textit{Intent}, and \textit{Difficulty Level}.
A sample of the generated query is in Table~\ref{table:query samples}. In the final phase, we harness the prompt illustrated in Figure~\ref{fig:prompt_filter_pe} and pass it through Deepseek-Coder-V2-Instruct~\cite{ds-v2} that filters out the queries based on their quality.
These configurations ensure that the generated data reflects diverse development scenarios, resulting in high-fidelity training data that closely mirrors real-world developer interactions.

\begin{figure}[!htbp]
    \centering
    \includegraphics[width=\columnwidth]{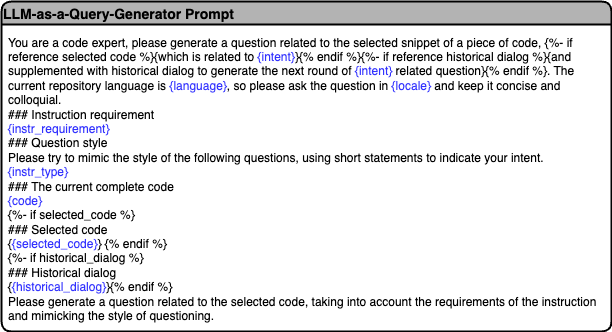}
    \caption{Prompt used by the query-generating.}
    \label{fig:query generating prompt}
    \vspace{-10pt}
\end{figure}

\begin{figure}[!htbp]
    \centering
    \includegraphics[width=\columnwidth]{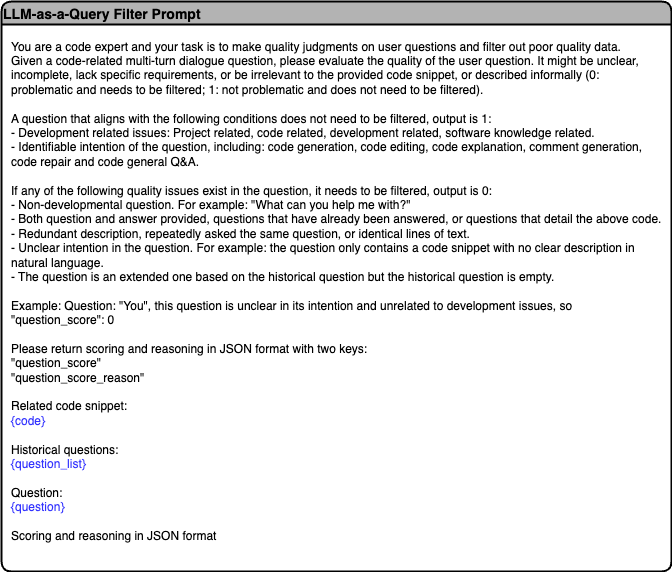}
    \caption{Prompt used by the query quality filter.}
    \label{fig:prompt_filter_pe}
    \vspace{-10pt}
\end{figure}

\begin{table*}[!htbp]
    \centering
    \caption{Samples of the Generated Query}
    \footnotesize{
    \begin{tabular}{|c|c|c|p{8cm}|}   
    \hline                       
    \textbf{Intent} & \textbf{Difficulty Level} & \textbf{Response Reference Regions} & \textbf{Query} \\   
    \hline
    Comment Generation & Elementary & Selected Code & Generate comments for the currently selected code. \\ 
    \hline
    Comment Generation & Intermediate & Historical Dialog & Any specific comments needed for the 'new' method in Rust to clarify its purpose and usage? \\ 
    \hline
    Code Editing & Advanced & Context & Add new code logic, xlsx\_file\_path is set as a multi-line string, get the path and then loop through xlsx\_to\_txt for each path. \\ 
    \hline
    Code Editing & Expert & Selected Code & Please refactor this code to optimize it for best performance. \\ 
    \hline
    Code Explanation & Elementary & Selected Code & Can you explain the significance of ‘runApp(MyApp())’ in the context of this app's structure? \\ 
    \hline
    Code Explanation & Intermediate & Historical Dialog, & Can you go into more detail? \\ 
    \hline
    Code Repair & Advanced & Question & init(android.os.Parcel) failed to verify. What's the problem? \\
    \hline
    \multirow{3}*{Code Repair} & \multirow{3}*{Expert} & \multirow{3}*{Error Messages} & Query: Please fix the code. \\
    & & & Error Messages in prompt: Code: use rand::Rng; \\
    & & & Error Messages: unresolved import 'rand'; use of undeclared crate or module 'rand' \\
    \hline
    Code Generation & Elementary & General Knowledge & Write a bubbling sort algorithm. \\
    \hline
    Code Generation & Intermediate & Historical Dialog & Continue to generate subsequent. \\
    \hline
    Code General Q\&A & Advanced & Question & Using python to solve the equation dy/dx=5y+2x. \\
    \hline
    Code General Q\&A & Expert & General Knowledge & Make a python drawing software, list the libraries you use, give demos. \\
    \hline
    \end{tabular}
    }
    \label{table:query samples}
\end{table*}

\subsection{UI Automation in the IDE Plugin}

Due to the complexity of replicating real IDE environments and the lack of a low-cost solution to operate the IDE via APIs, we employ a UI automation tool to simulate user interactions within the IDE.
This tool executes predictable tasks such as selecting files, positioning cursors, and triggering the Q\&A interface, based on the chat configurations.
The Q\&A triggering process is shown in Figure~\ref{fig:Interact}. 
According to the input chat configurations, the UI automation tool will first pull down the repository and open the target file, the cursor position will first determine whether the code is selected or not, and then determine whether to move the cursor to the target position.
In the next step, the tool will trigger the online Q\&A process by clicking on the icon to get to the Q\&A interface and entering the query in the dialog box.
It captures the resulting Q\&A interactions, which are stored and iteratively refined for further use in model training.
The use of UI automation not only ensures data diversity but also dramatically reduces the manual effort required for generating large-scale datasets.
Details of the UI automation process are shown in pseudo code in Algorithm~\ref{alg:UI automation process}.


\begin{figure*}[!htbp]
    \centering
    \includegraphics[width=0.8\textwidth]{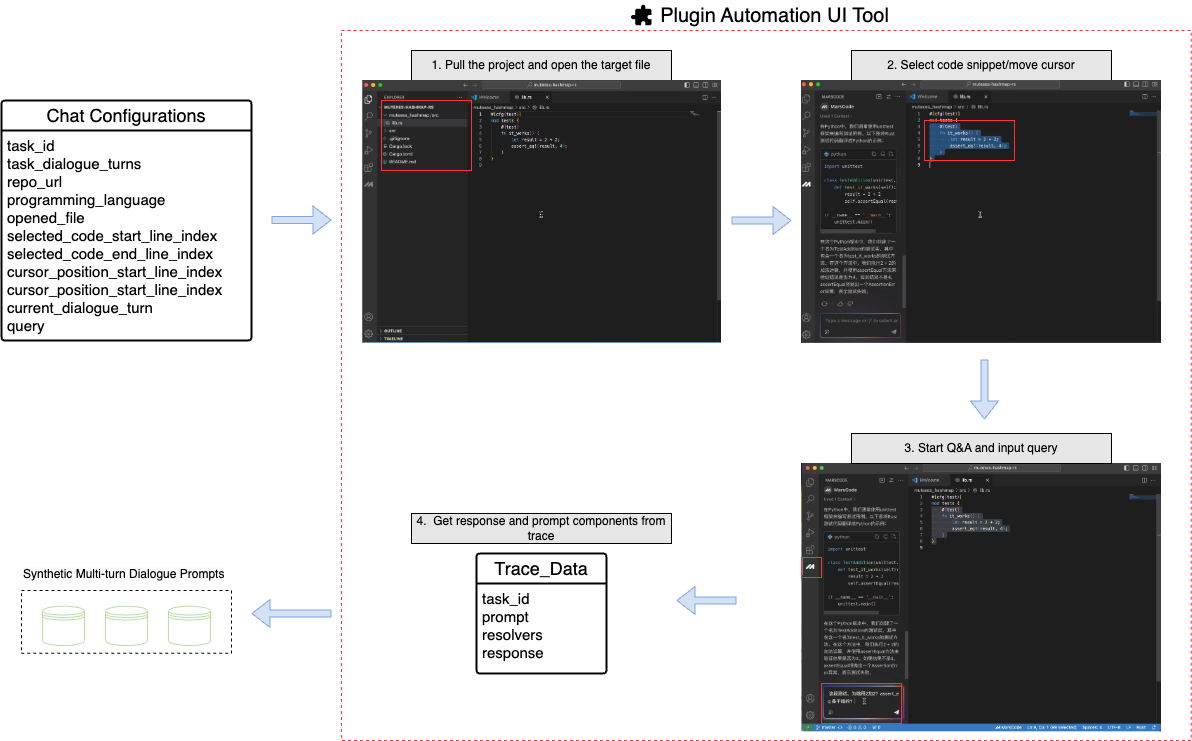}
    \caption{Framework of the Automation UI Tool}
    \label{fig:Interact}
\end{figure*}

\begin{algorithm}
    \renewcommand{\algorithmicrequire}{\textbf{Input:}}
    \renewcommand{\algorithmicensure}{\textbf{Output:}}
    \caption{Plugin UI automation}
    \label{alg:UI automation process}
    \footnotesize{
    \begin{algorithmic}[1]
        \REQUIRE Chat Configurations
        \STATE Pull the target code repository;
        \STATE Open the target file;
        \STATE $task\_dialogue\_turns \gets$ number of rounds of dialogue;
        \FOR {${t}\,$ in $\,task\_dialogue\_turns$}
            \IF {$selected\_code\_start\_line\_index$ \\  \textbf{and} $selected\_code\_end\_line\_index$}
                \STATE Selected the target code;
            \ELSIF {$cursor\_position\_start\_line\_index$ \\ \textbf{and} $cursor\_position\_end\_line\_index$}
                \STATE Move the cursor to the target position;
            \ENDIF 
            \STATE Open the marscode Q\&A interaction page;
            \STATE Type in a query to trigger an online Q\&A process;
            \STATE Capture the necessary information from the trace file;
            \STATE Store the Q\&A information in the database;
        \ENDFOR
        \ENSURE Multi-turn Dialogue Prompts
    \end{algorithmic}
    }
\end{algorithm}

\subsection{Response Generator}

After generating diverse and complex queries, high-quality responses are crucial for producing a useful training dataset.
To achieve this efficiently, we employ a multi-model response generation pipeline using models from the DeepSeek series~\cite{ds-v25, ds-v2}.
The pipeline comprises two phases: \textbf{response generation} and \textbf{response judgment}.

\subsubsection{Response Generation}
In the response generation process, we select three LLMs from the same series as our response generators. These include Deepseek-Coder-33B-Instruct\cite{ds-33b}, Deepseek-Coder-V2-Instruct\cite{ds-v2} and Deepseek-V2.5\cite{ds-v25}, all of which perform well on code benchmarks such as HumanEval~\cite{chen2021evaluating} and EvalPlus~\cite{liu2024your}.
Each query produced by UI automation is used to prompt all models to generate responses, forming a set of candidate responses for training. In the next phase, scoring will be conducted to select the best response for each query.

However, not every query has a perfect response as sometimes all candidate responses are not of high quality.
Instructions without an high-quality response will be filtered out, which may lead to the waste of instructions.
Therefore, we also optimize the pool of LLMs to enhance the probability of generating perfect responses. 
This is achieved through a self-improvement method, that is, fine-tuning the models within the pool using the data synthesized from \toolname{} itself.

\subsubsection{Response Judgment}

In the judgment process, we use the LLM-as-a-Judge methodology\cite{judge} to annotate candidate responses with GPT-4o.
We combine LLM scoring, ranking and rule-based deduction to select the optimal answer for each query:

\begin{itemize}
\item \textbf{Response scoring.} We use GPT-4o to score each answer from 1 to 5, comprehensively evaluating the ability to follow instructions and the quality of the answer. GPT-4o is asked to first output the rationale, and then produce the score. The prompt used is given in Figure~\ref{fig:response_score_prompt}.
\item \textbf{Rule-based deduction.} For response requirements that can be easily judged by rules but where the model has a chance of making incorrect judgments, we employ a rule-based deduction method for scoring. This score will be subtracted from the score obtained in the Response Scoring phase to determine the final score of the answer. Response that ultimately score 5-point will be used as our final training set. The deduction rules are detailed in Table~\ref{table:deduction rules}. 
\item \textbf{Response comparison.} If an instruction has more than one 5-point response, we will use GPT-4o to rank the responses and select the best one for the training set. The prompt used for comparison is given in Figure~\ref{fig:response_comparison_template}.
\end{itemize}

\begin{figure*}
    \centering
    \includegraphics[width=0.7\textwidth]{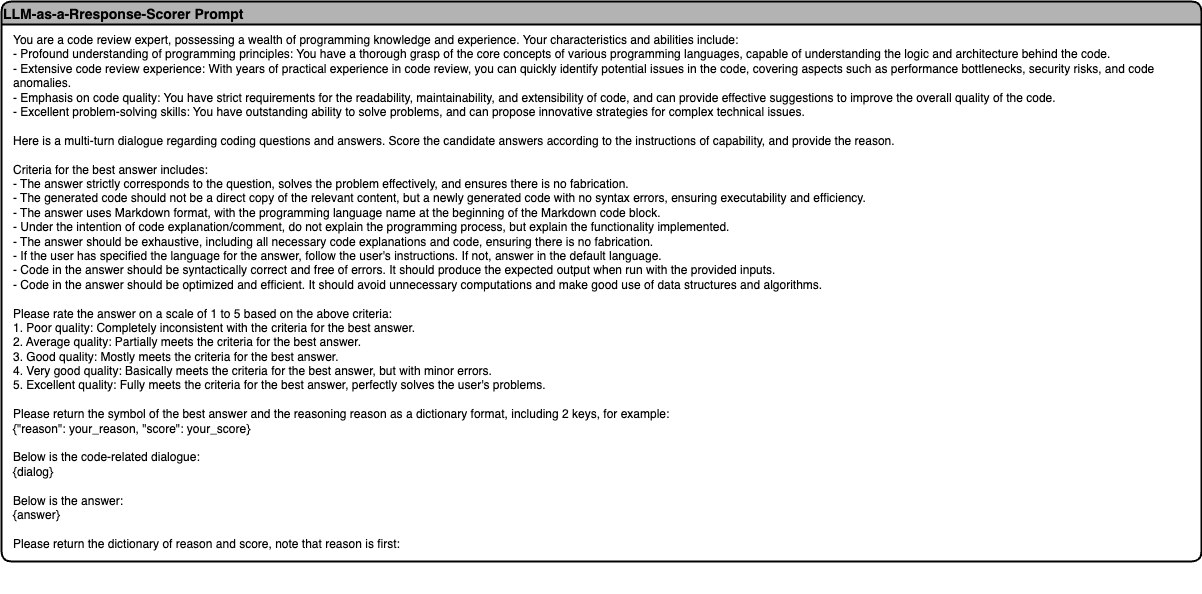}
    \caption{Response Scoring Template}
    \label{fig:response_score_prompt}
\end{figure*}

\begin{figure*}
    \centering
    \includegraphics[width=0.7\textwidth]{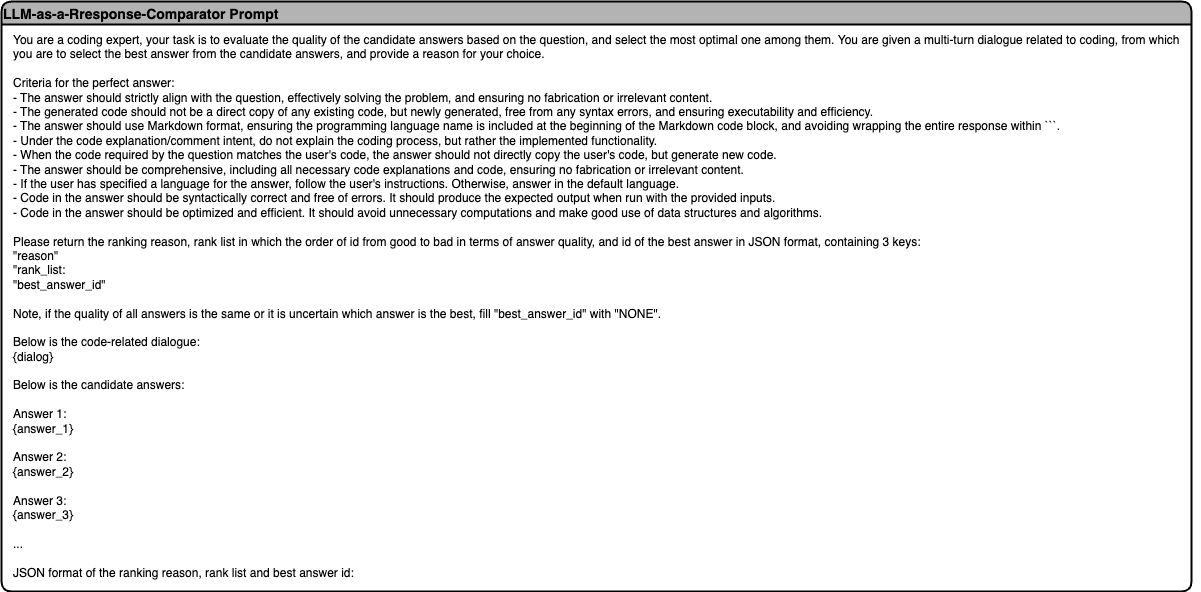}
    \caption{Response Comparison Template}
    \label{fig:response_comparison_template}
\end{figure*}

We only select top-scoring answers for training, and queries without a 5-point candidate answer will not be included in the training set to ensure the quality of the training data.

\begin{table*}[ht]
    \centering
    \caption{The Deduction Rules}
    \footnotesize{
    \begin{tabular}{|c|c|c|}   
    \hline                       
    
    \textbf{Scene} & \textbf{Deduction Item} & \textbf{Deduction Score} \\   
    \hline
    Inline Chat & Text description before the code &  1 \\    
    \hline
    Chat View & Lack of basic text description  &  1 \\  
    \hline
    \multirow{3}{*}{Inline Chat \& Chat View} 
    
    & Language of response inconsistent with the instruction request and system setting. &  1 \\  
    \cline{2-3}
 
    & Incomplete code markdown symbols &  1 \\
    \cline{2-3}
    & Altering the original code when editing is no required & 2 \\  
    \cline{2-3}
    & Revealing the requirements in the prompt & 2 \\
    \cline{2-3}
    & Incomplete response, truncated in the middle of words or code & 5 \\
    \cline{2-3}
    \hline
    \end{tabular}
    }
    \label{table:deduction rules}
\end{table*}


\section{Experiment}
\label{sec:experiment}
\begin{figure}
    \centering
    \includegraphics[scale=0.10]{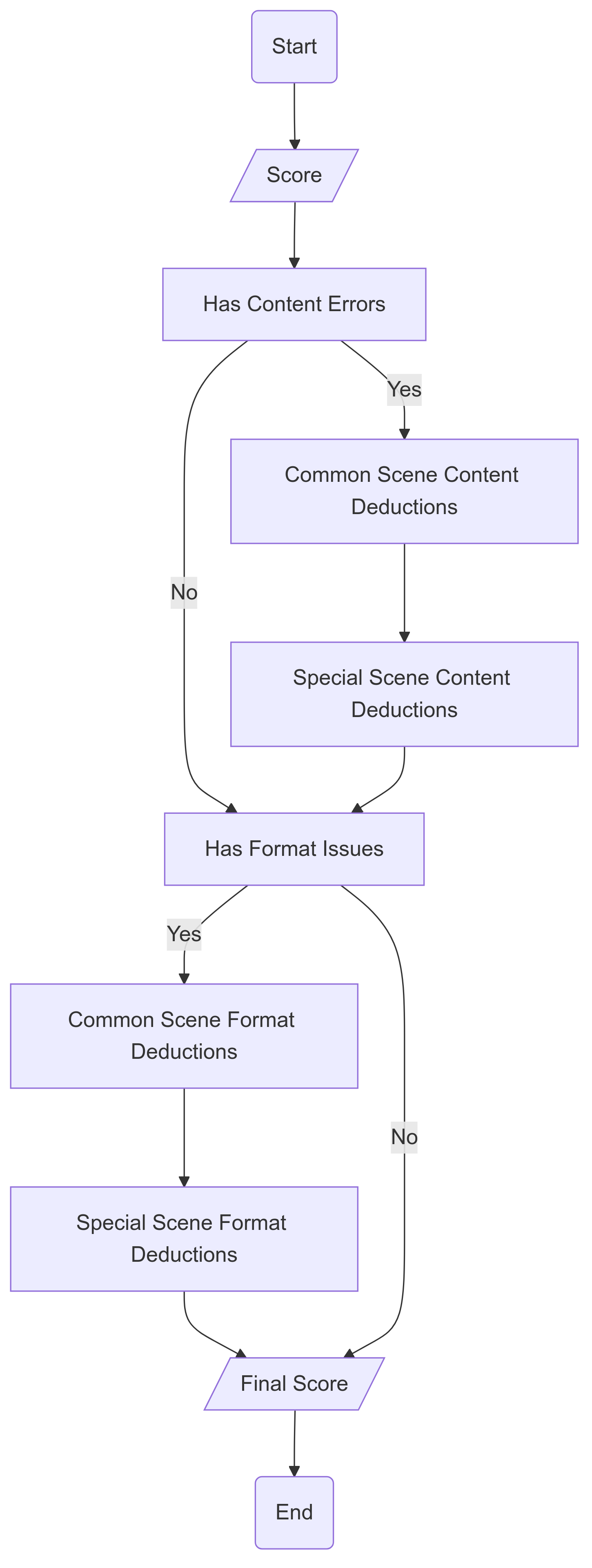}
    \caption{Process of Scoring}
    \label{fig:Scoring Process}
\end{figure}

\begin{table*}[!htbp]
    \centering
    \caption{The Penalty Points System}
    \footnotesize{
    \begin{tabular}{|c|c|p{10cm}|}   
    \hline                       
    \textbf{Category} & \textbf{Intent} & \textbf{Penalty Points} \\   
    \hline
    \multirow{5}*{Content Errors} & Common Scene & code errors, description error, response repetition, programming language errors, response truncation, not meeting user requirements \\
    \cline{2-3}
    & \multirow{4}*{Special Scene} & Code Explanation: nonsense content, missing critical content \\
    & & Comment Generation: nonsense content, missing function header comments, function header comments out of specification, comment locale error \\
    & & Code Repair: missing/redundant code repair \\
    & & Code Editing: missing/redundant code editing \\
    \hline
    \multirow{4}*{Format Issues} & Common Scene & locale error, markdown error, response formatting issues \\
    \cline{2-3}
    & \multirow{3}*{Special Scene} & Code Repair: lack of complete fix code, lack of basic text description \\
    & & Inline Chat: text description before the code \\
    & & Chat View: lack of basic text description \\
    \hline
    \end{tabular}
    }
    \label{table:penalty points system}
\end{table*}

\begin{table}[!htbp]
    \centering
    \caption{Manual Scoring Results}
    \footnotesize{
    \begin{tabular}{ccc}
    \hline
    \textbf{Model} & ${\textbf{PSR}}$ & ${\textbf{UR}}$ \\
    \cline{1-3}
    \textbf{DS-33B-Inst} & 26.00\% & 39.00\% \\
    \textbf{DS-33B-Inst-SFT(Ours)} & \textcolor{red}{$\uparrow$}\textcolor{red}{46.64\%} & \textcolor{red}{$\uparrow$}\textcolor{red}{59.40\%} \\
    \hline
    \end{tabular}
    }
    \label{table:manual scoring results}
\end{table}

\begin{table*}[!htbp]
    \centering
    \caption{Manual Scoring Results in Code Generation Intents}
    \footnotesize{
    \begin{tabular}{ccc|cc|cc}
    \hline
    \multirow{2}*{\textbf{Model}} & \multicolumn{2}{c|}{\textbf{Code Repair}} & \multicolumn{2}{c|}{\textbf{Code Generation}} & \multicolumn{2}{c}{\textbf{Code Editing}} \\
    \cline{2-7}
    & ${\textbf{PSR}}$ & ${\textbf{UR}}$ & ${\textbf{PSR}}$ & ${\textbf{UR}}$ & ${\textbf{PSR}}$ & ${\textbf{UR}}$ \\
    \hline
    \textbf{DS-33B-Inst} & 19.23\% & 44.44\% & 28.13\% & 53.13\% & 22.50\% & 32.50\% \\
    \textbf{DS-33B-Inst-SFT(Ours)} & \textcolor{red}{$\uparrow$}\textcolor{red}{44.44\%} & \textcolor{red}{$\uparrow$}\textcolor{red}{55.56\%} & \textcolor{red}{$\uparrow$}\textcolor{red}{56.25\%} & \textcolor{red}{$\uparrow$}\textcolor{red}{72.00\%} & \textcolor{red}{$\uparrow$}\textcolor{red}{65.00\%} & \textcolor{red}{$\uparrow$}\textcolor{red}{82.50\%} \\
    \hline
    \end{tabular}
    }
    \label{table:manual scoring results in code generation intens}
\end{table*}

\begin{table*}[!htbp]
    \centering
    \caption{Manual Scoring Results in Code Comprehension Intents}
    \footnotesize{
    \begin{tabular}{ccc|cc|cc}
    \hline
    \multirow{2}*{\textbf{Model}} & \multicolumn{2}{c|}{\textbf{Code Explanation}} & \multicolumn{2}{c|}{\textbf{Code General Q\&A}} & \multicolumn{2}{c}{\textbf{Comment Generation}} \\
    \cline{2-7}
    & ${\textbf{PSR}}$ & ${\textbf{UR}}$ & ${\textbf{PSR}}$ & ${\textbf{UR}}$ & ${\textbf{PSR}}$ & ${\textbf{UR}}$ \\
    \hline
    \textbf{DS-33B-Inst} & 19.23\% & 38.46\% & 44.07\% & 67.78\% & 16.67\% & 23.33\% \\
    \textbf{DS-33B-Inst-SFT(Ours)} & \textcolor{red}{$\uparrow$}\textcolor{red}{40.46\%} & \textcolor{red}{$\uparrow$}\textcolor{red}{53.85\%} & \textcolor{red}{$\uparrow$}\textcolor{red}{50.74\%} & \textcolor{red}{$\uparrow$}\textcolor{red}{69.26\%} & \textcolor{red}{$\uparrow$}\textcolor{red}{78.57\%} & \textcolor{red}{$\uparrow$}\textcolor{red}{78.57\%} \\
    \hline
    \end{tabular}
    }
    \label{table:manual scoring results in code comprehension intens}
\end{table*}

\begin{table}[!htbp]
    \centering
    \caption{Manual Scoring Results for Different Generated Data}
    \footnotesize{
    \begin{tabular}{ccc}
    \hline
    \textbf{Model} & ${\textbf{PSR}}$ & ${\textbf{UR}}$ \\
    \cline{1-3}
    \textbf{DS-33B-Inst-SFT(Ours)} & \textcolor{red}{46.64\%} & \textcolor{red}{59.40\%} \\
    \textbf{DS-33B-Inst-SFT(w/o MTD)} & 43.10\% & 52.86\% \\
    \textbf{DS-33B-Inst-SFT(subset)} & 46.28\% & 58.11\% \\
    \textbf{DS-33B-Inst-SFT(subset w/o MTD)} & 42.47\% & 52.18\% \\
    \textbf{DS-33B-Inst-SFT(Random)} & 41.47\% & 53.54\% \\
    
    \hline
    \end{tabular}
    }
    \label{table:data type vs}
\end{table}

\begin{table*}[!htbp]
    \centering
    \caption{Manual Scoring Results for Different Generated Data in Code Generation Intents}
    \footnotesize{
    \begin{tabular}{ccc|cc|cc}
    \hline
    \multirow{2}*{\textbf{Model}} & \multicolumn{2}{c|}{\textbf{Code Repair}} & \multicolumn{2}{c|}{\textbf{Code Generation}} & \multicolumn{2}{c}{\textbf{Code Editing}} \\
    \cline{2-7}
    & ${\textbf{PSR}}$ & ${\textbf{UR}}$ & ${\textbf{PSR}}$ & ${\textbf{UR}}$ & ${\textbf{PSR}}$ & ${\textbf{UR}}$ \\
    \hline
    \textbf{DS-33B-Inst-SFT(Ours)} & 44.44\% & 55.56\% & 56.25\% & \textcolor{red}{72.00\%} & \textcolor{red}{65.00\%} & \textcolor{red}{82.50\%} \\
    \textbf{DS-33B-Inst-SFT(w/o MTD)} & 42.78\% & 58.33\% & 53.13\% & 65.63\% & 61.54\% & 69.23\% \\
    \textbf{DS-33B-Inst-SFT(subset)} & \textcolor{red}{51.11\%} & \textcolor{red}{61.11\%} & \textcolor{red}{62.50\%} & 71.88\% & 55.00\% & 72.50\% \\
    \textbf{DS-33B-Inst-SFT(subset w/o MTD)} & 42.78\% & 52.78\% & 53.13\% & 59.38\% & 55.00\% & 64.50\% \\
    \textbf{DS-33B-Inst-SFT(Random)} & 42.78\% & 52.78\% & 50.00\% & 71.88\% & 60.00\% & 62.50\% \\
    
    \hline
    \end{tabular}
    }
    \label{table:data type vs in code generation intens}
\end{table*}

\begin{table*}[!htbp]
    \centering
    \caption{Manual Scoring Results for Different Generated Data in Code Comprehension Intents}
    \footnotesize{
    \begin{tabular}{ccc|cc|cc}
    \hline
    \multirow{2}*{\textbf{Model}} & \multicolumn{2}{c|}{\textbf{Code Explanation}} & \multicolumn{2}{c|}{\textbf{Code General Q\&A}} & \multicolumn{2}{c}{\textbf{Comment Generation}} \\
    \cline{2-7}
    & ${\textbf{PSR}}$ & ${\textbf{UR}}$ & ${\textbf{PSR}}$ & ${\textbf{UR}}$ & ${\textbf{PSR}}$ & ${\textbf{UR}}$ \\
    \hline
    \textbf{DS-33B-Inst-SFT(Ours)} & 40.46\% & 53.85\% & 50.74\% & \textcolor{red}{69.26\%} & \textcolor{red}{78.57\%} & \textcolor{red}{78.57\%} \\
    \textbf{DS-33B-Inst-SFT(w/o MTD)} & \textcolor{red}{50.00\%} & 53.85\% & \textcolor{red}{59.26\%} & 62.96\% & 53.33\% & 63.33\% \\
    \textbf{DS-33B-Inst-SFT(subset)} & 46.15\% & \textcolor{red}{65.38\%} & \textcolor{red}{59.26\%} & 62.96\% & 55.17\% & 68.97\% \\
    \textbf{DS-33B-Inst-SFT(subset w/o MTD)} & \textcolor{red}{50.00\%} & 61.54\% & 48.15\% & 55.56\% & 46.67\% & 63.33\% \\
    \textbf{DS-33B-Inst-SFT(Random)} & 33.33\% & 50.00\% & 57.69\% & 65.38\% & 50.00\% & 66.67\% \\
    \hline
    \end{tabular}
    }
    \label{table:data type vs in code comprehension intens}
\end{table*}

\begin{table}[!htbp]
    \centering
    \caption{Manual Scoring Results in New Evaluation Data}
    \footnotesize{
    \begin{tabular}{ccc}
    \hline
    \textbf{Model} & ${\textbf{PSR}}$ & ${\textbf{UR}}$ \\
    \cline{1-3}
    \textbf{DS-33B-Inst-SFT(subset)} & 57.91\% & 63.92\% \\
    \textbf{DS-33B-Inst-SFT(Ours)} & \textcolor{red}{$\uparrow$}\textcolor{red}{61.90\%} & \textcolor{red}{$\uparrow$}\textcolor{red}{64.17\%} \\
    \hline
    \end{tabular}
    }
    \label{table:data type vs-new evaluation data}
\end{table}

\begin{table*}[!htbp]
    \centering
    \caption{Manual Scoring Results in Code Generation Intents for New Evaluation Data}
    \footnotesize{
    \begin{tabular}{ccc|cc|cc}
    \hline
    \multirow{2}*{\textbf{Model}} & \multicolumn{2}{c|}{\textbf{Code Repair}} & \multicolumn{2}{c|}{\textbf{Code Generation}} & \multicolumn{2}{c}{\textbf{Code Editing}} \\
    \cline{2-7}
    & ${\textbf{PSR}}$ & ${\textbf{UR}}$ & ${\textbf{PSR}}$ & ${\textbf{UR}}$ & ${\textbf{PSR}}$ & ${\textbf{UR}}$ \\
    \hline
    \textbf{DS-33B-Inst-SFT(subset)} & 46.00\% & 48.00\% & 80.00\% & 80.00\% & 48.00\% & 60.00\% \\
    \textbf{DS-33B-Inst-SFT(Ours)} & \textcolor{red}{$\uparrow$}\textcolor{red}{50.00\%} & \textcolor{red}{$\uparrow$}\textcolor{red}{50.00\%} & \textcolor{red}{$\uparrow$}\textcolor{red}{82.00\%} & \textcolor{red}{$\uparrow$}\textcolor{red}{84.00\%} & \textcolor{red}{$\uparrow$}\textcolor{red}{56.00\%} & \textcolor{red}{$\uparrow$}\textcolor{red}{62.00\%} \\
    \hline
    \end{tabular}
    }
    \label{table:data type vs in code generation intens-new evaluation data}
\end{table*}

\begin{table*}[!htbp]
    \centering
    \caption{Manual Scoring Results in Code Comprehension Intents for New Evaluation Data}
    \footnotesize{
    \begin{tabular}{ccc|cc|cc}
    \hline
    \multirow{2}*{\textbf{Model}} & \multicolumn{2}{c|}{\textbf{Code Explanation}} & \multicolumn{2}{c|}{\textbf{Code General Q\&A}} & \multicolumn{2}{c}{\textbf{Comment Generation}} \\
    \cline{2-7}
    & ${\textbf{PSR}}$ & ${\textbf{UR}}$ & ${\textbf{PSR}}$ & ${\textbf{UR}}$ & ${\textbf{PSR}}$ & ${\textbf{UR}}$ \\
    \hline
    \textbf{DS-33B-Inst-SFT(subset)} & 66.00\% & 72.00\% & 54.00\% & 64.00\% & 56.00\% & 64.00\% \\
    \textbf{DS-33B-Inst-SFT(Ours)} & \textcolor{red}{$\uparrow$}\textcolor{red}{78.00\%} & \textcolor{red}{$\uparrow$}\textcolor{red}{78.00\%} & {54.00\%} & \textcolor{red}{$\uparrow$}\textcolor{red}{68.00\%} & {56.00\%} & {64.00\%} \\
    \hline
    \end{tabular}
    }
    \label{table:data type vs in code comprehension intens-new evaluation data}
\end{table*}

\begin{table*}[!htbp]
    \centering
    \caption{Manual Scoring Results for Response Generators with Optimization}
    \footnotesize{
    \begin{tabular}{c|c|c|c|c|c|c}
    \hline
    \multirow{1}*{\textbf{Model}} & \multicolumn{1}{c|}{\textbf{Code Repair}} & \multicolumn{1}{c|}{\textbf{Code Generation}} & \multicolumn{1}{c|}{\textbf{Code Editing}} & \multicolumn{1}{c|}{\textbf{Code Explanation}} & \multicolumn{1}{c|}{\textbf{Code General Q\&A}} & \multicolumn{1}{c}{\textbf{Comment Generation}} \\
    \cline{2-7}
    & ${\textbf{PSR}}$ & ${\textbf{PSR}}$ & ${\textbf{PSR}}$ & ${\textbf{PSR}}$ & ${\textbf{PSR}}$ & ${\textbf{PSR}}$ \\
    \hline
    \textbf{LLM-Pool} & 72.72\% & 89.65\% & 86.48\% & 94.18\% & 92.10\% & 52.27\% \\
    \textbf{LLM-Pool-SFT} & \textcolor{red}{$\uparrow$}\textcolor{red}{85.71\%} & \textcolor{red}{$\uparrow$}\textcolor{red}{96.55\%} & \textcolor{red}{$\uparrow$}\textcolor{red}{91.89\%} & \textcolor{red}{$\uparrow$}\textcolor{red}{98.84\%} & \textcolor{red}{$\uparrow$}\textcolor{red}{94.74\%} & \textcolor{red}{$\uparrow$}\textcolor{red}{71.59\%} \\
    \hline
    \end{tabular}
    }
    \label{table:Manual Scoring Results for Response Generators}
\end{table*}

\begin{table}[!htbp]
    \centering
    \caption{Manual Scoring Results for Response Generators}
    \footnotesize{
    \begin{tabular}{ccc}
    \hline
    \textbf{Model} & ${\textbf{PSR}}$ \\
    \cline{1-2}
    \textbf{LLM Pool (all models)} & \textcolor{red}{79.76\%} \\
    \textbf{DS-33B-Inst} & 26.60\% \\
    \textbf{DS-V2.5} & 64.65\% \\
    \textbf{DS-Coder-V2-Inst} & 68.14\% \\
    
    \hline
    \end{tabular}
    }
    \label{table:Manual Scoring Results for Pool of LLMs}
\end{table}

We investigate the following research questions (RQ) to study the effectiveness and efficiency of \toolname{}:

\begin{description}
    \item[RQ1.] What is the impact of efficiently expanding high-quality code question-and-answer data on code Q\&A systems?
    \item[RQ2.] How can synthetic data maintain consistency with real development environments, particularly in multi-turn dialogue data based on historical conversations?
    \item[RQ3.] What is the quality of automatically generated code question-and-answer responses?
\end{description}

We discuss evaluation metrics, results and analysis for each research question in the rest of this section.

\subsection{Experiment Setup}

We first de-identified real developer Q\&A data to ensure compliance with privacy standards.
For evaluation, we balanced the dataset according to user intent and programming languages, selecting a random sample of 300 entries.
The evaluation process was carried out by developers proficient in various programming stacks who manually scored the model’s responses.

The scoring process is shown in Figure~\ref{fig:Scoring Process}. 
Each model’s final score is the average of two rounds of inference, with each round going through three stages: initial scoring, quality inspection, and final score confirmation.
A team of 21 annotators handled the scoring process: 13 for initial scoring, 6 for secondary quality inspection, and 2 for final score confirmation.
We used a penalty points system (Table~\ref{table:penalty points system}) for scoring, with a maximum score of 5.

We established two key evaluation metrics:
\begin{itemize}
    \item \textbf{Usability Rate (UR):} The proportion of responses deemed usable ($Final Score \geq 4$).
    \item \textbf{Perfect Score Rate (PSR):} The proportion of responses judged as perfect ($Final Score = 5$).
\end{itemize}


We use Deepseek-coder-33B-Instruct as the seed model and employ high-quality Q\&A data produced by \toolname{} for supervised finetuning (SFT) , and each SFT model iterates for 3 epochs and follows the experimental setup of~\cite{ds-33b}.
During inference, we generated two response variations per prompt using a temperature of 0.3 and top-p of 0.95. The results were averaged over two rounds of inference.

\subsection{RQ1. Impact of High-Quality Code Q\&A Data on Code Q\&A Systems}

In this section, we assess the impact of \toolname{} on six key code Q\&A tasks: code generation, code editing, code explanation, comment generation, code repair, and general Q\&A.
The comparison results between the seed model (DeepSeek-33B-Instuct) and the fine-tuned model (DeepSeek-33B-Instruct-SFT) are shown in Tables~\ref{table:manual scoring results}, ~\ref{table:manual scoring results in code generation intens} and ~\ref{table:manual scoring results in code comprehension intens}.

We observe significant improvements across all tasks and metrics for the fine-tuned model. For code generation tasks such as code repair, code generation, and code editing, UR increased by 11\%, 19\%, and 50\%, respectively. For code understanding tasks such as code explanation, general Q\&A, and comment generation, UR improved by 15\%, 2\%, and 55\%, respectively.


To study the efficiency compared with human annotators, we asked the annotation team to construct code Q\&A pairs with the help of online search and other intelligent tools.
However, manual construction of code Q\&A pairs is time-consuming, yielding only 30 pairs per day per annotator.
In contrast, \toolname{} can generate 1,440 pairs per day per instance — 4.8 times the productivity of human annotators.
By scaling computational resources, multiple automation instances can further increase production capacity. Additionally, human inspection also reported that synthetic data generated by \toolname{} demonstrates superior variability and consistency with real-world distributions compared to manually generated data.


\subsection{RQ2. Consistency with the Development Environment}

Ww designed 3 groups of ablation experiments to verify the effectiveness of its online consistency, data type increment, and data increment, respectively, and the overall experimental results are compared as in Table~\ref{table:data type vs}, the results of the code generation intention as in Table~\ref{table:data type vs in code generation intens}, and the results of the code understanding intention as in Table~\ref{table:data type vs in code comprehension intens}.

\textbf{Synthetic data's consistency with online reality}. 
We generated a batch of data randomly, incorporating elements such as arbitrary programming languages, selected code fragments, and queries, etc. By employing identical training and inference parameters, we fine-tune the Deepseek-coder-33B-Instruct based on the two types of synthetic data, separately. The comparison result is DS-33B-Inst-SFT(Random) VS DS-33B-Inst-SFT(Ours). Evidently, the data generated by \toolname{} overall performs better, an absolute increase of ${5\%}$ in ${PSR}$ and ${6\%}$ in ${UR}$. It's noteworthy that in code editing and comment generation, the absolute values of ${UR}$ have improved by more than ${10\%}$.

\textbf{Data type increment of synthetic data}. Multi-turn dialogue capability of a conversational model is important for the performance of IDE code Q\&A.
However, in the complex task of generating high-quality multi-turn dialogue data, our \toolname{} provides robust support. We compared the effectiveness of fine-tuned models using synthetic data, considering the presence or absence of multi-turn dialogue data (abbreviated as MTD), i.e., DS-33B-Inst-SFT(Ours) VS DS-33B-Inst-SFT(w/o MTD) and DS-33B-Inst-SFT(subset) VS DS-33B-Inst-SFT(subset w/o MTD). The results indicate that the multi-turn dialogue data generated by \toolname{} can further enhance the performance of the SFT model, with an absolute increase of ${3\%}$ in the ${PSR}$ and ${6\%}$ in the ${UR}$. For intents with a stronger historical dialogue dependency (code editing and code generation), there is at least a ${7\%}$ improvement in ${UR}$.

\textbf{Increment of synthetic data}. DS-33B-Inst-SFT(subset) is a subset of the final version dataset DS-33B-Inst-SFT(Ours) with half the data volume. The comparison results indicate that the model with the increased SFT data from \toolname{} can sustain stable effects on the old period evaluation set.

In the same time, the final version of DS-33B-Inst-SFT(Ours) also includes a new batch of data generated based on online data from another time period, and we sampled online data from this time period to serve as a new evaluation data. The comparison results in the new evaluation data are presented in Table~\ref{table:data type vs-new evaluation data}, ~\ref{table:data type vs in code generation intens-new evaluation data} and ~\ref{table:data type vs in code comprehension intens-new evaluation data}. Notably, the model with the increased SFT data generated in the new period performs significantly better on the new period's online data, with ${PSR}$ achieving a ${4\%}$ increase, ${UR}$ a ${1\%}$ increase.

Given that \toolname{} can autonomously generate data closely resembling online user behavior based on developers' online activity, we can infer that an SFT model based on \toolname{} can achieve continuous automatic optimization.

\subsection{RQ3. Automating High-Quality Code Q\&A Responses}
To ensure the quality of the generated responses, we evaluate the overall quality and optimization benefits of the response generator, and also assessed the consistency between the automated judgment system and human scoring.

Our annotation team manually score 430 data samples with responses generated by each LLM in our response generator pool.
Table \ref{table:Manual Scoring Results for Pool of LLMs} shows the overall $PSR$ of the LLM pool and the $PSR$ of each LLM in the pool.
Although the PSR of each model is not high, when combined together, it reaches 79.76\%. Since each model excels in different types of questions, \textbf{the approach of using the LLM pool as a response generator combines the strengths of the various LLMs}, enabling 79.76\% of the data produced in one round has the quality to be included in the training set. Despite this, some prompts still lack perfect answers and will be added to the next production chain until a perfect answer appears. This has to some extent affects the efficiency of our data production. In order to increase the proportion of data that can be used as a training set in each round of data production, we also iteratively optimize the LLM pool through self-improvement method. Table \ref{table:Manual Scoring Results for Response Generators} illustrates that after fine-tuning all the models in the LLM pool with the training set from the last production round, the ${PSR}$ for each intent significantly increases and the average ${PSR}$ has an increase of 8.65\%. The ${PSR}$ for the Comment Generation intent, particularly, increased by 19.32\%.

To evaluate the accuracy of our automated judgment system, our annotation team manually scores 600 responses with distinct instructions. Since the responses we ultimately select for the training set are only 5-point responses, the accuracy rate of scoring 5-point responses is used as a metric to assess our judement system:
\begin{equation}
Accuracy_5 = \frac{pred_{5}\textunderscore pos}{pred_{5}\textunderscore pos + pred_{5}\textunderscore neg}
\end{equation}

$Accuracy_5$ represents the proportion of 5-point responses scored by our judgement system that align with the scores given by human annotators. It measures the degree of agreement between the judgement system's scoring of 5-point responses and the human evaluations. Also, ${pred_{5}\textunderscore pos}$ indicates instances where both judgement system and human annotation assigned a score of 5, and ${pred_{5}\textunderscore neg}$ represents instances where the judgement system assigned a score of 5, but the human annotation did not. 

Considered human scoring as the ground truth, the $Accuracy_5$ for the judgement system is 88.47\%. Among the ${pred_{5}\textunderscore neg}$ instances, 96.00\% are usable responses, scored more than 3-point by human annotators. 81.62\% of the responses manually scored as 5-point are recalled by the system. This indicates that our judgment system has a relatively high consistency with human scoring, and most of the inconsistently scored 5-point responses are still usable.

\section{Discussion and Lessons Learned}
\label{sec:discussion}

\textbf{Industrial deployment confirms the practicality of \toolname.}
\toolname{} is integrated into the model training of our in-house programming assistant, MarsCode, which offers functionalities such as code completion, intelligent Q\&A, code generation, explanation, and repair. Along with the launch of \toolname's SFT model, the online effectiveness indicator-acceptance rate (percentage of shown answer accepted by the user) was improved by \textbf{33\%}.
The deployment of \toolname{} has demonstrated its practicality and effectiveness in real-world code Q\&A scenarios, and also marks the beginning of a continuous improvement cycle. \toolname{}’s success in industrial application highlights its role in advancing AI-powered development tools.

\textbf{Insights from Online User Behavior}
Through analyzing real-world user interactions, several key behavioral patterns have been identified, which provide valuable insights for further optimizing \toolname.

\begin{itemize}
    \item \textbf{Cursor Behavior:} The top 3 cursor behaviors observed among users are: \textit{no active file (40\%)}, \textit{having an active file (33\%)} and \textit{selecting a code block (35\%)}. These behaviors indicate that users often asking questions in a passive browsing state, which highlights the \textbf{necessity of automated context retrieval to locate files of interest} to help models better understand the context.
    \item \textbf{Instruction Type:} Although pre-configured quick chat buttons are provided, such as \textit{explain code} and \textit{generate comments}, which are configured with pre-defined prompt templates and will appear when a code snippet is selected by the user, two-thirds of users prefer to directly input their questions instead of using these chat options. This finding highlights the need for natural and flexible query handling in the Q\&A tool.
    \item \textbf{Dialog Turns:} Most conversations between users and the assistant consist of 5 turns or fewer, with a near $1:1$ ratio of single-turn to multi-turn dialogs. To improve the effectiveness of multi-turn dialogs, it is essential to first enhance the quality of single-turn responses, ensuring that users find value in the tool and continue engaging with it over longer conversations.
    \item \textbf{Response Reference Regions:} When responding to users, the most referenced information sources are the selected code, the user's question, and general programming knowledge. This suggests that focusing on these reference origins can significantly improve the relevance and accuracy of the responses provided by the tool.
    \item \textbf{User Intent Classification:} After classifying the intent of user questions, each type of intent has a dominant subcategory, which accounts for nearly half of the interactions. For example:
        \begin{itemize}
            \item Code Explanation: Function and method explanations are the most frequent.
            \item Code Generation: Test code generation is the most common.
            \item General Q\&A: Usage of programming language syntax is the most prevalent query.
            \item Code Repair: The most triggered is fixing compilation errors.
            \item Code Editing: Code translation to another programming language is the most common editing request.
            \item Comment Generation: Comments for functions and methods are the most frequent requests.
        \end{itemize}
\end{itemize}

\textbf{The Limitation of \toolname{}}. Despite the undeniable advantages of \toolname{}, there are certain limitations that need to be addressed. More scenarios expansion requires some development labor. Although it can efficiently handle mainstream code Q\&A scenarios, customization and adaptation for more niche or emergent application areas pose a challenge, demanding further manpower and expertise. Additionally, there's a need for a finer classification of scenarios, such as classifications based on knowledge points, e.g., code editing including code optimization, code refactoring, code translation, etc. Such intricate classifications can enrich the diversity of the synthetic data and improve the SFT effect of synthetic data on LLM. Moreover, we will continually broaden and optimize our model pool to keep pace with the rapid development of LLM and meet the ever-evolving requirements of developers. The optimization of the model pool can effectively improve the recall rate of high-quality responses, which in turn greatly improves the robustness and versatility of \toolname{}. This ongoing optimization process underscores our commitment to delivering a tool that stays relevant in the face of dynamic development landscapes and user expectations.

\section{Conclusion and Future Work}
\label{sec:conclusion}
In this paper, we explored the challenges and solutions associated with producing high-quality synthetic training data in the scenario of developing LLM-based programming assistants, while ensuring data security and privacy. We introduced \toolname{} capable of efficiently generating synthetic data that mimics real coding scenarios. This tool synthesizes data based on various user perspectives, significantly improving the performance of fine-tuned models in the empirical evaluation.
The deployment of \toolname{} to generate training data of our in-house model also experienced a 33\% of the improvement of the acceptance rate by our users.


In the future, we consider methods for synthesizing personalized responses and explore more data synthesis schemes and their effects during the reinforcement learning and direct preference optimization phases of model training.

\bibliographystyle{IEEEtran}
\bibliography{references}

\begin{thebibliography}{10}
\providecommand{\url}[1]{#1}
\csname url@samestyle\endcsname
\providecommand{\newblock}{\relax}
\providecommand{\bibinfo}[2]{#2}
\providecommand{\BIBentrySTDinterwordspacing}{\spaceskip=0pt\relax}
\providecommand{\BIBentryALTinterwordstretchfactor}{4}
\providecommand{\BIBentryALTinterwordspacing}{\spaceskip=\fontdimen2\font plus
\BIBentryALTinterwordstretchfactor\fontdimen3\font minus \fontdimen4\font\relax}
\providecommand{\BIBforeignlanguage}[2]{{%
\expandafter\ifx\csname l@#1\endcsname\relax
\typeout{** WARNING: IEEEtran.bst: No hyphenation pattern has been}%
\typeout{** loaded for the language `#1'. Using the pattern for}%
\typeout{** the default language instead.}%
\else
\language=\csname l@#1\endcsname
\fi
#2}}
\providecommand{\BIBdecl}{\relax}
\BIBdecl

\bibitem{cursor}
AnySphere, ``Cursor,'' \url{https://www.cursor.com/}, 2024.

\bibitem{copilot}
``Github copilot,'' \url{https://github.com/features/copilot}, accessed: 2024-05-28.

\bibitem{marscode}
``Marscode,'' \url{https://www.marscode.com/}, accessed: 2024-05-28.

\bibitem{codeium}
``Codeium,'' \url{https://codeium.com/}, accessed: 2024-05-28.

\bibitem{liu2024marscode}
Y.~Liu, P.~Gao, X.~Wang, C.~Peng, and Z.~Zhang, ``Marscode agent: Ai-native automated bug fixing,'' \emph{arXiv preprint arXiv:2409.00899}, 2024.

\bibitem{zhang2024autocoderover}
Y.~Zhang, H.~Ruan, Z.~Fan, and A.~Roychoudhury, ``Autocoderover: Autonomous program improvement,'' in \emph{Proceedings of the 33rd ACM SIGSOFT International Symposium on Software Testing and Analysis}, 2024, pp. 1592--1604.

\bibitem{xia2024agentless}
C.~S. Xia, Y.~Deng, S.~Dunn, and L.~Zhang, ``Agentless: Demystifying llm-based software engineering agents,'' \emph{arXiv preprint arXiv:2407.01489}, 2024.

\bibitem{macneil2023experiences}
S.~MacNeil, A.~Tran, A.~Hellas, J.~Kim, S.~Sarsa, P.~Denny, S.~Bernstein, and J.~Leinonen, ``Experiences from using code explanations generated by large language models in a web software development e-book,'' in \emph{Proceedings of the 54th ACM Technical Symposium on Computer Science Education V. 1}, 2023, pp. 931--937.

\bibitem{dong2023abilities}
G.~Dong, H.~Yuan, K.~Lu, C.~Li, M.~Xue, D.~Liu, W.~Wang, Z.~Yuan, C.~Zhou, and J.~Zhou, ``How abilities in large language models are affected by supervised fine-tuning data composition,'' \emph{arXiv preprint arXiv:2310.05492}, 2023.

\bibitem{tian2022learning}
Z.~Tian, J.~Chen, Q.~Zhu, J.~Yang, and L.~Zhang, ``Learning to construct better mutation faults,'' in \emph{Proceedings of the 37th IEEE/ACM International Conference on Automated Software Engineering}, 2022, pp. 1--13.

\bibitem{feng2020codebert}
Z.~Feng, D.~Guo, D.~Tang, N.~Duan, X.~Feng, M.~Gong, L.~Shou, B.~Qin, T.~Liu, D.~Jiang \emph{et~al.}, ``Codebert: A pre-trained model for programming and natural languages,'' \emph{arXiv preprint arXiv:2002.08155}, 2020.

\bibitem{nijkamp2022codegen}
E.~Nijkamp, B.~Pang, H.~Hayashi, L.~Tu, H.~Wang, Y.~Zhou, S.~Savarese, and C.~Xiong, ``Codegen: An open large language model for code with multi-turn program synthesis,'' \emph{arXiv preprint arXiv:2203.13474}, 2022.

\bibitem{fried2022incoder}
D.~Fried, A.~Aghajanyan, J.~Lin, S.~Wang, E.~Wallace, F.~Shi, R.~Zhong, W.-t. Yih, L.~Zettlemoyer, and M.~Lewis, ``Incoder: A generative model for code infilling and synthesis,'' \emph{arXiv preprint arXiv:2204.05999}, 2022.

\bibitem{chen2021evaluating}
M.~Chen, J.~Tworek, H.~Jun, Q.~Yuan, H.~P. D.~O. Pinto, J.~Kaplan, H.~Edwards, Y.~Burda, N.~Joseph, G.~Brockman \emph{et~al.}, ``Evaluating large language models trained on code,'' \emph{arXiv preprint arXiv:2107.03374}, 2021.

\bibitem{wei2024magicoder}
Y.~Wei, Z.~Wang, J.~Liu, Y.~Ding, and L.~Zhang, ``Magicoder: Empowering code generation with oss-instruct,'' in \emph{Forty-first International Conference on Machine Learning}, 2024.

\bibitem{liu2024mftcoder}
B.~Liu, C.~Chen, Z.~Gong, C.~Liao, H.~Wang, Z.~Lei, M.~Liang, D.~Chen, M.~Shen, H.~Zhou \emph{et~al.}, ``Mftcoder: Boosting code llms with multitask fine-tuning,'' in \emph{Proceedings of the 30th ACM SIGKDD Conference on Knowledge Discovery and Data Mining}, 2024, pp. 5430--5441.

\bibitem{tan2024large}
Z.~Tan, A.~Beigi, S.~Wang, R.~Guo, A.~Bhattacharjee, B.~Jiang, M.~Karami, J.~Li, L.~Cheng, and H.~Liu, ``Large language models for data annotation: A survey,'' \emph{arXiv preprint arXiv:2402.13446}, 2024.

\bibitem{park2024leveraging}
J.~Park, P.~Wisniewski, and V.~Singh, ``Leveraging large language models (llms) to support collaborative human-ai online risk data annotation,'' \emph{arXiv preprint arXiv:2404.07926}, 2024.

\bibitem{zendel2024enhancing}
O.~Zendel, J.~S. Culpepper, F.~Scholer, and P.~Thomas, ``Enhancing human annotation: Leveraging large language models and efficient batch processing,'' in \emph{Proceedings of the 2024 Conference on Human Information Interaction and Retrieval}, 2024, pp. 340--345.

\bibitem{bhat2023large}
S.~Bhat and V.~Varma, ``Large language models as annotators: A preliminary evaluation for annotating low-resource language content,'' in \emph{Proceedings of the 4th Workshop on Evaluation and Comparison of NLP Systems}, 2023, pp. 100--107.

\bibitem{yu2024fine}
Y.~Yu, G.~Rong, H.~Shen, H.~Zhang, D.~Shao, M.~Wang, Z.~Wei, Y.~Xu, and J.~Wang, ``Fine-tuning large language models to improve accuracy and comprehensibility of automated code review,'' \emph{ACM Transactions on Software Engineering and Methodology}, 2024.

\bibitem{ouyang2022training}
L.~Ouyang, J.~Wu, X.~Jiang, D.~Almeida, C.~Wainwright, P.~Mishkin, C.~Zhang, S.~Agarwal, K.~Slama, A.~Ray \emph{et~al.}, ``Training language models to follow instructions with human feedback,'' \emph{Advances in neural information processing systems}, vol.~35, pp. 27\,730--27\,744, 2022.

\bibitem{ds-33b}
D.~Guo, Q.~Zhu, D.~Yang, Z.~Xie, K.~Dong, W.~Zhang, G.~Chen, X.~Bi, Y.~Wu, Y.~Li, F.~Luo, Y.~Xiong, and W.~Liang, ``Deepseek-coder: When the large language model meets programming -the rise of code intelligence,'' \emph{arXiv preprint arXiv:2401.14196}, 2024.

\bibitem{ds-v2}
DeepSeek-AI, Q.~Zhu, D.~Guo, Z.~Shao, D.~Yang, P.~Wang, R.~Xu, Y.~Wu, Y.~Li, H.~Gao, S.~Ma, W.~Zeng \emph{et~al.}, ``Deepseek-coder-v2: Breaking the barrier of closed-source models in code intelligence,'' \emph{arXiv preprint arXiv:2406.11931}, 2024.

\bibitem{lozhkov2024starcoder}
A.~Lozhkov, R.~Li, L.~B. Allal, F.~Cassano, J.~Lamy-Poirier, N.~Tazi, A.~Tang, D.~Pykhtar, J.~Liu, Y.~Wei, T.~Liu, M.~Tian, D.~Kocetkov, A.~Zucker, Y.~Belkada, Z.~Wang, Q.~Liu, D.~Abulkhanov, I.~Paul, Z.~Li, W.-D. Li, M.~Risdal, J.~Li, J.~Zhu, T.~Y. Zhuo, E.~Zheltonozhskii, N.~O.~O. Dade, W.~Yu, L.~Krauß, N.~Jain, Y.~Su, X.~He, M.~Dey, E.~Abati, Y.~Chai, N.~Muennighoff, X.~Tang, M.~Oblokulov, C.~Akiki, M.~Marone, C.~Mou, M.~Mishra, A.~Gu, B.~Hui, T.~Dao, A.~Zebaze, O.~Dehaene, N.~Patry, C.~Xu, J.~McAuley, H.~Hu, T.~Scholak, S.~Paquet, J.~Robinson, C.~J. Anderson, N.~Chapados, M.~Patwary, N.~Tajbakhsh, Y.~Jernite, C.~M. Ferrandis, L.~Zhang, S.~Hughes, T.~Wolf, A.~Guha, L.~von Werra, and H.~de~Vries, ``Starcoder 2 and the stack v2: The next generation,'' 2024.

\bibitem{ds-v25}
DeepSeek-AI, A.~Liu, B.~Feng, B.~Wang, B.~Wang, B.~Liu, C.~Zhao, C.~Dengr, C.~Ruan, D.~Dai, D.~Guo, D.~Yang, D.~Chen, D.~Ji, E.~Li, F.~Lin \emph{et~al.}, ``Deepseek-v2: A strong, economical, and efficient mixture-of-experts language model,'' \emph{arXiv preprint arXiv:2405.04434}, 2024.

\bibitem{liu2024your}
J.~Liu, C.~S. Xia, Y.~Wang, and L.~Zhang, ``Is your code generated by chatgpt really correct? rigorous evaluation of large language models for code generation,'' \emph{Advances in Neural Information Processing Systems}, vol.~36, 2024.

\bibitem{judge}
L.~Zheng, W.-L. Chiang, Y.~Sheng, S.~Zhuang, Z.~Wu, Y.~Zhuang, Z.~Lin, Z.~Li, D.~Li, E.~P. Xing, H.~Zhang, J.~E. Gonzalez, and I.~Stoica, ``Judging llm-as-a-judge with mt-bench and chatbot arena,'' \emph{arXiv preprint arXiv:2306.05685}, 2023.

\end{thebibliography}

\end{document}